# Novel Highly Active Pt/Graphene Catalyst for Cathodes of Cu(II/I)-Mediated Dye-Sensitized Solar Cells

Ladislav Kavan[1,2*], Hana Krysova[1], Pavel Janda[1], Hana Tarabkova[1], Yasemin Saygili[3], Marina Freitag[3,4], Shaik M. Zakeeruddin[2], Anders Hagfeldt[3] and Michael Grätzel[2]

[1]*J. Heyrovsky Institute of Physical Chemistry, v.v.i., Academy of Sciences of the Czech Republic, Dolejskova 3, CZ-18223 Prague 8, Czech Republic*
[2]*Laboratory of Photonics and Interfaces, Institute of Chemical Sciences and Engineering, Swiss Federal Institute of Technology, CH-1015 Lausanne, Switzerland*
[3]*Laboratory of Photomolecular Science, Institute of Chemical Sciences and Engineering, Swiss Federal Institute of Technology, CH-1015, Lausanne, Switzerland*
[4]*Department of Chemistry, Ångström Laboratory, Uppsala University, 751 20 Uppsala, Sweden*

*e-mail: kavan@jh-inst.cas.cz

**Abstract.** Novel highly active, optically-transparent electrode catalyst containing Pt, PtO$_x$, graphene oxide and stacked graphene platelet nanofibers is developed for a cathode of Cu(II/I)-mediated dye-sensitized solar cells. The catalyst layer is deposited on a FTO substrate, which thus becomes smoother than the parent FTO, but the button-like Pt/PtO$_x$ nanoparticles are still distinguishable. The found electrocatalytic activity for the Cu(tmby)$_2{}^{2+/+}$ redox couple (tmby is 4,4',6,6'-tetramethyl-2,2'-bipyridine) is outperforming that of alternative catalysts, such as PEDOT or platinum. Exchange current densities exceeding 20 mA/cm$^2$ are provided exclusively by our novel catalyst. The synergic boosting of electrocatalytic activity is seen, if we normalize it to the catalytic performance of individual components, i.e. Pt and graphene nanofibers. The outstanding properties of our cathode are reflected by the performance of the corresponding solar cells using the Y123-sensitized titania photoanode. Champion solar-conversion efficiency (11.3 % at 0.1 sun) together with a fill factor of 0.783 compare favorably to all other so far reported best values for this kind of solar cells and the given experimental conditions.

# 1. Introduction

The dye sensitized solar cell (DSC) is popular for high efficiency, low-cost, and amazing versatility of its construction [1–3]. Current progress in DSCs is highlighted by replacement of the traditional redox mediator (i.e. $I_3^-/I^-$) with other high-voltage redox couples such as Co(III/II) [4–14] or Cu(II/I) complexes [15–20] which provide larger open-circuit voltage, $V_{OC}$, of solar cells. The Co(III/II) redox couples showed fast charge-transfer at graphene electrodes [9–14]. This finding was behind all the recent efficiency benchmarks of DSCs: 12% [6], 13% [7] up to the current record (14.3%) [8].

There has been a quite conflicting debate about the electrocatalytic activity of carbons [9,21–23], but Roy-Mayhew et al. [24] demonstrated recently that the activity of graphene for the Co(III/II) redox couple comes actually from the presence of lattice defects (dangling bonds at lattice vacancies and edges), rather than from oxygen-containing surface groups, as it was hypothesized by others. The cited work reported on 250-times greater electrocatalytic activity (referenced to that of HOPG being a surrogate of perfect graphene) for low-oxygen graphene but rich with such defects [24].

Other redox system, offering even larger $V_{OC}$ values are based on tetra-coordinated Cu(II/I) complexes with distorted tetragonal symmetry [15–20]. The solar conversion efficiency of Cu-mediated DSCs at AM 1.5 (1 sun) illumination intensity has developed from the initial value of 1.4% [25] to 7% [26], 8.3% [18], 10.3% [16,17] up to the recent record of 11.3% (achieved by using a combination of two judiciously designed dyes) [20]. The largest efficiencies were observed for $Cu(tmby)_2^{2+/+}(TFSI)_{2/1}$ where tmby is 4,4',6,6'-tetramethyl-2,2'-bipyridine, and TFSI is bis(trifluoromethylsufonyl)imide [16,20]. Similar efficiencies were observed for an analogous complex with 6,6'-dimethyl-2,2'-bipyridine and $BF_4^-$ counterion [17]. Still larger efficiency (13.2 %) was reported for the champion (two-dyes) DSC, but at smaller light intensity (0.12 sun illumination) [20]. The mentioned Cu-complexes have highly positive redox potentials, but still work surprisingly well with the Y123 dye (Fig. S1 Supp. Info) illustrating the fundamentally interesting fact that the driving force for dye regeneration can be as small as 0.1 eV in DSCs [16].

These Cu complexes were reported to show sluggish electrochemistry on certain noble metals, oxides and carbon black, unless the latter was platinized [26]. Earlier works on Cu-mediated DSCs employed Pt-coated FTO as the cathode (FTO is F-doped $SnO_2$). The Pt-catalyst was fabricated by the traditional synthetic protocol based on thermal decomposition of $H_2PtCl_6$ [27], by sputtering [25] or by ALD deposition of platinum inverse opal, the latter being considerably more active than the flat Pt electrode [27,28]. The most popular cathode catalyst was poly(3,4-ethylenedioxythiophene), PEDOT [15,16,18,19] which provided the best performing DSCs so far [16,20]. Recently, high-efficiency DSC (10.3 %) was also reported using a commercial Pt-counterelectrode [17].

The use of graphene as the cathode catalyst for Cu(II/I)-based redox mediators was first explored by ourselves in [15]. Here, we have chosen the so called "stacked graphene platelet nanofiber" (SGNF) from ABCR/Strem which turned out to be the top-performing catalyst also for various other Co-based mediators [7,14,29]. The charge-transfer resistances measured by electrochemical impedance spectroscopy were regularly better for the SGNF-coated FTO, compared to those of the Pt-coated FTO, but did not beat the values for PEDOT-coated FTO [15]. Furthermore, the adhesion of SGNF or other graphene-based catalysts to FTO is known to be poor [9–12,14,23] which is another handicap against PEDOT or Pt. The adhesion problem can be minimized by using a composite of graphene-based catalyst with graphene oxide instead of pure catalyst [10,14,15]. In this case, the graphene oxide acts as 'mortar' connecting the nanoparticles mutually and with the FTO substrate. The optimized material (called GONF80) was composed from 80 wt% of graphene platelet nanofibers and 20 wt% of graphene oxide; it turned out to be highly active for both the Co-based [14] and Cu-based [15] mediators.

In a search for still better catalyst, we have been inspired by the work of Cheng et al. [30] reporting a 'double-layer' cathode catalyst for the traditional I-mediated DSC. In this work [30] the few-layers graphene (made by CVD) was transferred to FTO and subsequently over-coated by Pt grown by sputtering. An alternative (and actually the better-performing architecture) was the reverse one: FTO was first coated by Pt and then by graphene [30]. Here, we have employed the same idea, but simplified the deposition protocol: Pt was made by thermal decomposition of $H_2PtCl_6$ and the graphene-based counterpart was GONF80 which was deposited from aqueous solution. Our results are essentially consistent with the earlier works on I-mediated DSCs, confirming that the combination of Pt with graphene [31–33] and other carbons [34] is beneficial for the counterelectrode performance. More specifically, our newly developed catalyst is outperforming all other earlier systems, i.e. Pt, pure SGNF, GONF80 and PEDOT. For the first time, we demonstrate here practical solar cells with Cu-mediators and graphene-based cathode catalysts. Though our DSCs do not beat the champion devices at 1 sun illumination, they provided the efficiency of 11.3 % at 0.1 sun which it the largest so far reported value for the Y123 dye and this light intensity (cf. [16]).

## 2. Experimental Section

Chemicals

The Cu(tmby)$_2$TFSI$_{2/1}$ complexes were available from our earlier work [16]. Briefly, the Cu(I) counterpart of the complex was synthesized by the reaction of CuI with tmby, and the Cu(II) counterpart was prepared by the reaction of CuCl$_2$ with tmby [16]. Other chemicals were from Aldrich or Merck, and used as received from the supplier.

Electrode materials

FTO glass (TEC 15 from Libbey-Owens-Ford, 15 Ohm/sq) was ultrasonically cleaned in a detergent solution (Deconex) followed by sonication in water, ethanol and acetone (30 min each). The PEDOT-coated FTO electrodes were fabricated by an electrochemical deposition from EDOT as detailed in [16]. Platinized FTO was prepared by deposition of 5 µL/cm$^2$ of 10 mM H$_2$PtCl$_6$ in 2-propanol and calcination at 400$^o$C for 15 minutes. Stacked graphene platelet nanofiber acid washed (SGNF) was from ABCR/Strem. According to the manufacturer's specification, this material (density 0.3 g/cm$^3$, surface area 120 m$^2$/g) is composed from fibers having a mean width of 40-50 nm and are 0.1 - 10 µm long. SGNF was dispersed in water by sonication and the solution was left overnight to separate big particles by sedimentation. The supernatant containing about 1 mg/mL was stable for several days without marked sedimentation. Single-layer graphene oxide (GO) was from Cheap Tubes, Inc. It was dispersed in water by sonication to a concentration of 1 mg/mL. The precursor for composite electrodes was prepared by mixing the SGNF dispersion with GO solution to a desired proportion of both components. The optimum concentration was 80 wt% of SGNF and 20 wt% of GO; this electrode is abbreviated GONF80 [14,15]. The graphene films on FTO were deposited from aqueous solutions either by drop-casting or by air-brush spraying. The amount of deposited carbon was quantified by measurement of optical density. Consistent with our previous works [11,12,14,35], the optical transmittance at a wavelength of 550 nm, $T_{550}$ served as a parameter characterizing the films. The final graphene-based films on FTO had the $T_{550}$ values between 70 and 85 %; PEDOT had ca. 60 % and Pt around 95 % transmittance.

The symmetrical dummy cell was fabricated from two identical FTO-supported electrodes which were separated by Surlyn (DuPont) tape as a seal and spacer. The sheet edges of FTO were coated by ultrasonic soldering (Cerasolzer alloy 246, MBR Electronics GmbH) to improve electrical contacts. The distance between electrodes was measured by a digital micrometer. The cell was filled with an electrolyte through two holes in the FTO support which was finally closed either by Kapton foil or by a Surlyn seal. Electrolyte solution was identical to that used in solar tests reported earlier [15,16]: 0.2 M Cu(tmby)$_2$TFSI + 0.04 M Cu(tmby)$_2$(TFSI)$_2$ + 0.1 M LiTFSI + 0.5 M *tert*-butylpyridine in acetonitrile.

Methods

Scanning electron microscopy (SEM) images were obtained by a Hitachi FE SEM S-4800 microscope. The optical spectra were measured by Perkin Elmer Lambda 1050 spectrometer with integrating sphere in transmission mode. The reference spectrum was air. Atomic force microscopy (AFM) images were obtained using Multimode Nanoscope IIIa (Bruker, USA) instrument. The following terms are used in description of surface morphology: "micrograin" (lateral diameter d ≥ 1µm), "sub-micrograin" ($10^2$ nm < d < 1 µm) and "nanograins" (d < 100 nm). Roughness parameters (Surface Area Difference, SAD and Roughness Average, Ra) were determined by Surface Roughness Analysis software Nanoscope III version 5.12r5 (Bruker).

X-ray photoelectron spectroscopy (XPS) was studied using Omicron Nanotechnology instrument equipped with a monochromatized AlKα source (1486.7 eV) and a hemispherical analyzer operating in constant analyzer energy mode with a multichannel detector. The CasaXPS program was used for spectra analysis.

Electrochemical measurements were carried out using Autolab PGstat-30 equipped with the FRA module (Ecochemie). Electrochemical impedance data were processed using Zplot/Zview software. The impedance spectra were acquired in the frequency range from 100 kHz to 0.1 Hz, at 0 V bias voltage, the modulation amplitude was 10 mV. Experiments on dye-sensitized solar cells were carried out as in our earlier work [15,16]. Briefly, FTO-supported, TiO$_2$ double-layer photoanode (5 µm nanocrystalline + 5 µm scattering layers) was prepared from Dyesol precursors and sensitized with Y123 dye (Dyenamo AB). The DSCs were assembled with 25 µm Surlyn (Dupont) spacer and sealant. The electrolyte solution was identical to that used for dummy cells (see above). The DSC's current-voltage characteristics were obtained by using a 450 W xenon light source (Osram XBO 450, Germany) with a filter Schott 113. The light power was regulated to the AM 1.5G solar standard by using a reference Si photodiode equipped with a color-matched filter (KG-3, Schott) to reduce the mismatch between the simulated light and AM 1.5G to less than 4% in the wavelength region of 350–750 nm. The differing intensities were regulated with neutral wire mesh attenuator. The applied potential and cell current were measured using a Keithley model 2400 digital source meter.

## 3. Results and Discussion

Our initial screening of various preparative protocols confirmed that the FTO/Pt/graphene catalyst was outperforming the alternative one with the opposite order of layers, i.e. FTO/graphene/Pt; this finding is consistent with the work of Cheng et al. [30]. Our optimized catalysts were made by using FTO with thermally deposited Pt as the parent substrate. This was subsequently over-coated with either pure SGNF (the catalyst is further abbreviated PtNF) or with a mixture of 80% of SGNF+20% of graphene oxide (further abbreviated PtGONF). While the PtNF catalyst was mechanically unstable (with easy delamination of SGNF particles from the substrate) the PtGONF was wear-resistant, in accord with earlier works on the graphene/graphene-oxide composites [10, 14, 15].

Figure 1 shows the AFM images of our starting substrates, i.e. a clean FTO (Fig. 1A), a FTO coated by platinum made from $H_2PtCl_6$ decomposition (Fig.1B) and a FTO coated by PtGONF catalyst (Fig. 1C). We clearly distinguish the difference between blank sub-micrograins of pure FTO and those covered by platinum nanograins and PtGONF. The overall surface roughness parameters Ra ≈ 15 nm and surface area difference, SAD ≈ 30%, determined from the images on Fig. 1 are similar for both clean and platinized FTO, but the surface coated by PtGONF is substantially smoother (Ra ≈ 11 nm, SAD ≈ 10%). This does not reflect the existence of platinum nanoparticles having size of about 20 nm in diameter and apex height 4-5 nm (Fig. S2, Supporting Info), which are overridden by dimensional fluctuations of FTO sub-micrograins reaching lateral dimensions typically orders of magnitude larger.

The corresponding SEM images are shown in Fig. 2. In general, the surface textures seen by AFM and SEM are reproduced. Though the smooth morphology of PtGONF catalyst is the dominating one, we occasionally detect also larger aggregates of graphene-based material on the surface of our catalyst. This is shown on AFM/SEM micrographs in Figs S3a and S3b (Supporting Info). Casual occurrence of these big particles is ascribed to polydisperse nature of our source material (SGNF), which contains fibers of varying lengths from 0.1 to 10 μm (see Experimental Section). Nevertheless, the biggest particles are obviously removed by sedimentation of the pre-sonicated solution during the preparation of the catalyst's precursor (see Experimental Section). AFM images of our catalyst (Figs 1 and S3A) indicate naked location of FTO grainy surface with dispersed Pt nanoparticles at the parent Pt@FTO surface (Fig. S4A), while areas covered by graphene-based composite show substantial smoothness with still visible Pt nanoparticles (Fig. 1C and S4B).

Wide scans X-ray photoelectron spectra (XPS) are shown in Fig. S5 at two takeoff angles of photoelectrons, i.e. $9^o$ and $90^o$; in the first case, the surface sensitivity is enhanced. Besides the omnipresent C1s signal, we occasionally detect impurities of Na, Ca and P in some samples in concentrations 0-4 at%. The F content is at the level of detection limit of our analysis (ca. 1 at%) even in a pure blank FTO. Detailed scans of the main photoelectron lines are shown in Fig. 3 (measured at the $90^o$ takeoff angle). The O1s binding energies increase in the order of samples: FTO < Pt@FTO < PtGONF@FTO, and certain binding-energy upshift is also observed for the Sn3d lines. The O1s envelope is broad in all cases, but for the sake of clarity only the O1s spectrum of the PtGONF sample is deconvoluted to the corresponding components. They obviously indicate a complicated array of chemical shifts (O-Sn-bulk, O-Sn-surface, O-C, O-Pt, O-H etc.) but we do not attempt any assignment at this stage of our research. Nevertheless, broadening of the O1s line toward higher binding energies was also observed by others for a nominally blank FTO, and was attributed to C-O bonds [36].

The deconvoluted signals of Pt $4f_{5/2}$ and $4f_{7/2}$ in our parent material (Pt@FTO) clearly confirm the presence of $PtO_x$ with Pt(II) and Pt(IV) components in addition to the main signal of metallic platinum, Pt(0). This conclusion is in accord with earlier studies of the Pt catalyst made by thermal decomposition of $H_2PtCl_6$ [37] and by other methods [33,34,38,39]. The final Pt/graphene composite catalyst (PtGONF) shows similar multicomponent features of $PtO_x$, and a specific C1s signal at 286.4 eV which is ascribed to C-O species of graphene oxide (arrow in Fig. 3). Very similar spectra were reported also for alternatively made catalyst of this kind, in which a mixture of $H_2PtCl_6$ and GO was sprayed from ethanolic solution on FTO [33]. A small C1s peak with larger binding energy was occasionally observed in blank FTO as well [36].

The electrocatalytic activity of DSCs' cathodes and mass transport in the electrolyte solution was studied using symmetrical dummy cells [9], in which the electrolyte solution with $Cu(tmby)_2^{2+/+}$ was sandwiched between two identical FTO-supported catalytic electrodes. Figure 4 presents cyclic voltammograms (left charts) and electrochemical impedance spectra (EIS, right charts) of these cells. The spectra were fitted to the equivalent circuit, model 1, Figure S6 (Supporting Info) of symmetrical dummy cells. Our data upgrade those in our earlier work, in which we had presented the spectra of simple catalysts, Pt, SGNF, PEDOT and GONF80 [15].

The inverse slope of cyclic voltammogram at the potential of 0 V characterizes the catalytic activity of an electrode; it is actually the overall cell resistance ($R_{CV}$) that can be attained at low current densities [11]. The found $R_{CV}$ (Fig. 4) is 29 $\Omega \cdot cm^2$ for PtGONF catalyst and 36 $\Omega \cdot cm^2$ for PtNF catalyst (Fig. 4). The limiting diffusion currents, $j_L$ are poorly distinguished in Fig. 4, but comparable (≈30 mA/cm$^2$) though they are also dependent on the dummy cell spacing ($\delta$):

$$j_L = 2nFcD/\delta \qquad (1)$$

($n = 1$ is the number of electrons, $F$ is the Faraday constant, $c$ is the concentration of diffusion-limited species, i.e. $Cu(tmby)_2^{2+}$ and $D$ is the diffusion coefficient). Assuming a spacing of $\delta \approx 30$ μm (defined by the Surlyn foil) then the current of 30 mA/cm$^2$ translates into $D \approx 1.2 \cdot 10^{-5}$ cm$^2$/s which is near the value of $2.2\ 10^{-5}$ cm$^2$/s reported earlier from rotating-disc experiments [16].

More detailed information about the electrocatalytic activity and mass transport in symmetrical dummy cell follows from EIS. The spectra were fitted to the model 1 equivalent circuit (Figure S6, Supporting Info) [15]. This model provided reasonably accurate fitting (Fig. 4, Table 1), though another more complicated circuit with two time-constants (model 2, Fig. S6) was recommended for analogous GONF80 electrodes (but Pt-free) [10,14,15]. Surprisingly, the spectrum of our still more complicated material (i.e. the triple-component catalyst, PtGONF) could be reasonably fitted to the simple model 1, and the same applies also for the PtNF catalyst (Fig. 4). In all cases, the catalytic activity is evaluated from the high-frequency part of the spectrum by a charge transfer resistance, $R_{CT}$, which is in series to the constant phase elements (CPE). The error of fitting of $R_{CT}$ values was ca. 6 % (PtGONF) and 12 % (PtNF). The use of CPE is necessary, due to the deviation from the ideal capacitance by the electrode inhomogeneity (roughness) [9]. The impedance of constant phase element equals:

$$Z_{CPE} = B^{-1} \cdot (i\omega)^{-\beta} \tag{3}$$

where ω is the frequency and $B$, $\beta$ are the frequency-independent parameters of the CPE ($0 \leq \beta \leq 1$; the corresponding parameters are 'CPE-T'= B and 'CPE-P = β).

The charge transfer at the counterelectrode can be alternatively evaluated from the high-frequency part of impedance spectra of complete DSC devices in dark at high applied forward bias (near $V_{oc}$) when the transport resistance in $TiO_2$ becomes negligible [1,2,14]. Figure 5 shows the corresponding spectrum of a solar cell with PtGONF cathode. Figure S7 (Supporting Info) shows comparative spectra of solar cells with PEDOT and Pt cathodes. The fitted $R_{CT}$ values are 0.80, 2.1 and 7.2 $\Omega cm^2$ for PtGONF, PEDOT and Pt, respectively (Table 1). The fitting error was between 3 and 8 %, i.e. comparable or smaller than that on symmetrical cells.

The fitted $R_{CT}$ values scale inversely with the exchange current density, $j_0$, at the cathode [9]:

$$j_0 = \frac{RT}{nFR_{CT}} = Fk_0(c_{ox}^{1-\alpha} \cdot c_{red}^{\alpha}) \tag{4}$$

$R$ is the gas constant, $T$ is temperature, $k_0$ is the formal (conditional) rate constant of the electrode reaction, $c_{ox}$ and $c_{red}$ are the concentrations of oxidized and reduced mediator, respectively and α is the charge-transfer coefficient (α ≈ 0.5). In good DSCs, the value of $j_0$ should be comparable to the short-circuit photocurrent density at 1 sun illumination. We note that our $R_{CT}$ found for PtGONF electrode (0.8 - 1.1 $\Omega \cdot cm^2$) is the best value among all other comparable ones for the Cu-mediated DSCs [15]. It translates (Eq. 4) into the exchange current of 23 to 32 $mA/cm^2$, which is well above the short-circuit current densities observed in our practical DSCs (Table 2, Fig. 6). Consequently, the charge-transfer rate at the counterelectrode is fast enough to support unperturbed function of our DSCs.

Our optimized catalyst (PtGONF) provides not only the impressive value of $j_0$ exceeding 20 $mA/cm^2$, but also the best value of serial resistance, $R_s$. Though it is tempting to ascribe this positive effect to the conductivity enhancement by graphene overlayer, the $R_s$ values are in general poorly defined, and depend on various other experimental parameters such as on the quality of soldered contacts to FTO. Another fundamentally interesting observation is the synergic quality improvement of PtGONF, referenced to that of the corresponding individual catalytic components, i.e. pure SGNF and Pt (Table 1). For instance the $j_0$ values are boosted by a factor of about 3 (normalized to SGNF) or 8 (normalized to Pt).

The low-frequency parts of the impedance spectra of both symmetrical cells and DSCs are dominated by the Warburg impedance. It is modelled by a finite-length element $W_s$ with the parameters 'W$_s$-R'=$R_W$, 'W$_s$-T'=$T_W$ and 'Ws-P' = 0.5 [40,41]. The finite length Warburg diffusion impedance is expressed as:

$$Z_W = \frac{R_W}{\sqrt{iT_W\omega}}\tanh\sqrt{iT_W\omega} \quad ; \quad T_W = \frac{\delta^2}{D} \tag{5}$$

The fitted values of symmetrical cells are collected in Table 1. The found diffusion resistances are all near 20 $\Omega \cdot cm^2$, which compare well to previously reported ones [15]. Obviously, the ionic transport in electrolyte solution should not be dependent on the electrode material, unless the latter is highly porous. Actually, the main complicating factor is the presence of *tert*-butylpyridine, which has a strong negative influence on both $R_w$ and $R_{CT}$. This problem still remains to be solved [15].

Finally, we tested our newly prepared counterelectrode catalyst (PtGONF) in dye-sensitized solar cells. Figure 6 presents example IV-curves for the Cu(tmby)$_2^{2+/+}$-mediator in acetonitrile solution. For comparison, we tested also a control device with the parent Pt@FTO counterelectrode. Figure S8 (Supporting Info) confirms that this electrode exhibits quite poor activity for the Cu(II/I)-redox mediator. This observation is interesting in view of the fact that the Pt@FTO is known to be highly active for alternative redox mediators (I-based and Co-based). The Cu(II/I)-couple is, indeed known to be quite sluggish on platinum, unless it is prepared in high-surface-area nanotexture (inverse opal) [27,28]. (Note that certain commercial Pt-electrodes can be reasonably active too [17]).

Table 2 summarizes detailed performance metrics of our DSCs. The remarkable photocurrent enhancement of the DSC with PtGONF counterelectrode is somewhat balanced by slightly smaller $V_{OC}$ of the same solar cell. Deeper inspection of Fig. 6 reveals larger dark currents for PtGONF, but surprisingly not for PtNF, where the delamination of carbon particles from the counterelectrode could occur [10]. At this stage of our research we do not have a solid explanation for this smaller $V_{OC}$. In terms of the classical model of DCS, the $V_{OC}$ is defined by a difference of the quasi-Fermi level of electrons in illuminated $TiO_2$ and the energy level equivalent to the redox mediator; hence, it should not depend on the counterelectrode (for further discussion see [10,11]).

The most promising performance data were observed for DSC with PtGONF cathode at 0.1 sun illumination. The efficiency of 11.3 % is the highest value ever reported for such experimental conditions, i.e. for the Y123 dye and 0.1 sun illumination [16]. Furthermore, the IV characteristic is close to perfection with fill factor of 0.783 (Fig. 6). This fill factor (Table 2) is very high not only among our set of data (Table 2) but also among all other so far reported best values for the Cu-mediated DSCs (0.78-0.79) [15,16,20].

**Conclusions**

A novel highly-active cathode catalyst was developed for the Cu(II/I)-mediated dyes-sensitized solar cells. The optimized catalyst (PtGONF) is a multi-component material, containing Pt, $PtO_x$, graphene oxide and stacked graphene platelet nanofibers. The catalyst is deposited on FTO substrate, which thus becomes smoother than the parent FTO, but the $Pt/PtO_x$ nanoparticles of lateral diameter ~20 nm and apex height 5 nm are still distinguishable in the final material. The catalytic film is semitransparent (with optical transmission of 70-85% at 550 nm wavelength) which is outperforming the transparency of PEDOT being a popular electrocatalyst for Cu-mediated DSCs.

The electrocatalytic activity of PtGONF for the $Cu(tmby)_2^{2+/+}$ redox couple in acetonitrile electrolyte solution (with charge-transfer resistance of 0.8 to 1.1 $\Omega.cm^2$) is the best one, compared to all other usual counterelectrode catalysts used in dye-sensitized solar cells of this type, i.e. PEDOT, Pt and pure graphene-based materials. Specifically this novel catalyst is capable of supporting the exchange current densities of >20 $mA/cm^2$ which are more than twice as big as those observed on the traditional catalyst, i.e. PEDOT. The enhancement is yet stronger if we normalize it to the performance of the individual catalytic components: the exchange currents are larger by a factor of about 3 (normalized to SGNF) or 8 (normalized to Pt). This evidences synergic quality improvement of PtGONF.

The good electrocatalytic activity of PtGONF is reflected also by the performance metrics of the corresponding solar cells using Y123 sensitized titania photoanode. Champion efficiency of 11.3 % was observed at 0.1 sun together with a fill factor of 0.783 which compares favorably to all other so far reported values for this kind of solar cells and experimental conditions.

**Acknowledgment**

This work was supported by the Grant Agency of the Czech Republic (contract No. 13-07724S), by the European Union H2020 Program (no. 696656–Graphene Flagship Core1) and by the Swiss National Science Foundation project "Fundamental studies of mesoscopic devices for solar energy conversion" with project number 200021_157135/1.

**Appendix A. Supplementary data**

Supplementary data associated with this article can be found, in the online version, at ….

**Table 1**. Electrochemical parameters of the studied cathode materials in symmetrical dummy cells obtained from fitting of impedance spectra.

| Electrode | $R_s$ ($\Omega.cm^2$) | $R_{CT}$[b] ($\Omega.cm^2$) | $j_0$[b] (mA/cm$^2$) | $R_W$ ($\Omega.cm^2.s^{-1/2}$) | CPE:B ($\Omega^{-1}.cm^{-2}.s^\beta$) | CPE: $\beta$ |
|---|---|---|---|---|---|---|
| PtGONF | 1.0 | 1.1 [0.80][b] | 23 [32][b] | 20.9 | 9.8·10$^{-3}$ | 0.79 |
| PtNF | 2.3 | 5.0 | 5.1 | 18.1 | 7.6·10$^{-3}$ | 0.78 |
| Pt[a] | 2.6 | 7.8 [7.2][b] | 3.3 [3.6][b] | 24.9 | 1.0·10$^{-5}$ | 0.89 |
| SGNF[a] | 2.9 | 2.9 | 8.9 | 21.7 | 3.1·10$^{-5}$ | 0.94 |
| PEDOT[a] | 2.4 | 3.0 [2.1][b] | 8.6 [12][b] | 22.7 | 3.6·10$^{-4}$ | 0.94 |

Notes:

a) Data from Ref. [15]

b) For comparison, the corresponding $R_{CT}$ and $j_0$ values from EIS on complete solar cells are quoted in brackets (see Figs. 5, S7 and the text below).

**Table 2.** Characteristics of solar cells with Y123-sensitized TiO$_2$ photoanodes at different illumination intensities. Short circuit photocurrent density = $j_{SC}$, open-circuit voltage = $V_{OC}$, fill factor = $FF$, solar conversion efficiency = $\eta$.

| Cathode | Illumination (sun) | $j_{SC}$ (mA/cm$^2$) | $V_{OC}$ (V) | $FF$ | $\eta$ (%) |
|---|---|---|---|---|---|
| PtGONF | 1 | 14.01 | 1.02 | 0.665 | 9.5 |
| PtGONF | 0.5 | 7.11 | 0.98 | 0.746 | 10.4 |
| PtGONF | 0.1 | 1.60 | 0.90 | 0.783 | 11.3 |
| PtNF | 1 | 1.30 | 1.05 | 0.661 | 9.1 |
| PtNF | 0.5 | 6.22 | 1.02 | 0.721 | 9.8 |
| PtNF | 0.1 | 11.90 | 0.92 | 0.743 | 9.7 |
| PEDOT | 1 | 13.32 | 1.05 | 0.637 | 8.9 |
| PEDOT | 0.5 | 6.72 | 1.02 | 0.704 | 9.7 |
| PEDOT | 0.1 | 1.47 | 0.93 | 0.755 | 10.4 |
| Pt@FTO[a] | 1 | 10.99 | 1.05 | 0.313 | 3.7 |
| Pt@FTO[a] | 0.5 | 5.82 | 1.02 | 0.422 | 4.9 |
| Pt@FTO[a] | 0.1 | 1.11 | 0.94 | 0.675 | 7.1 |

[a] Parent substrate made by thermal decomposition of H$_2$PtCl$_6$ at FTO

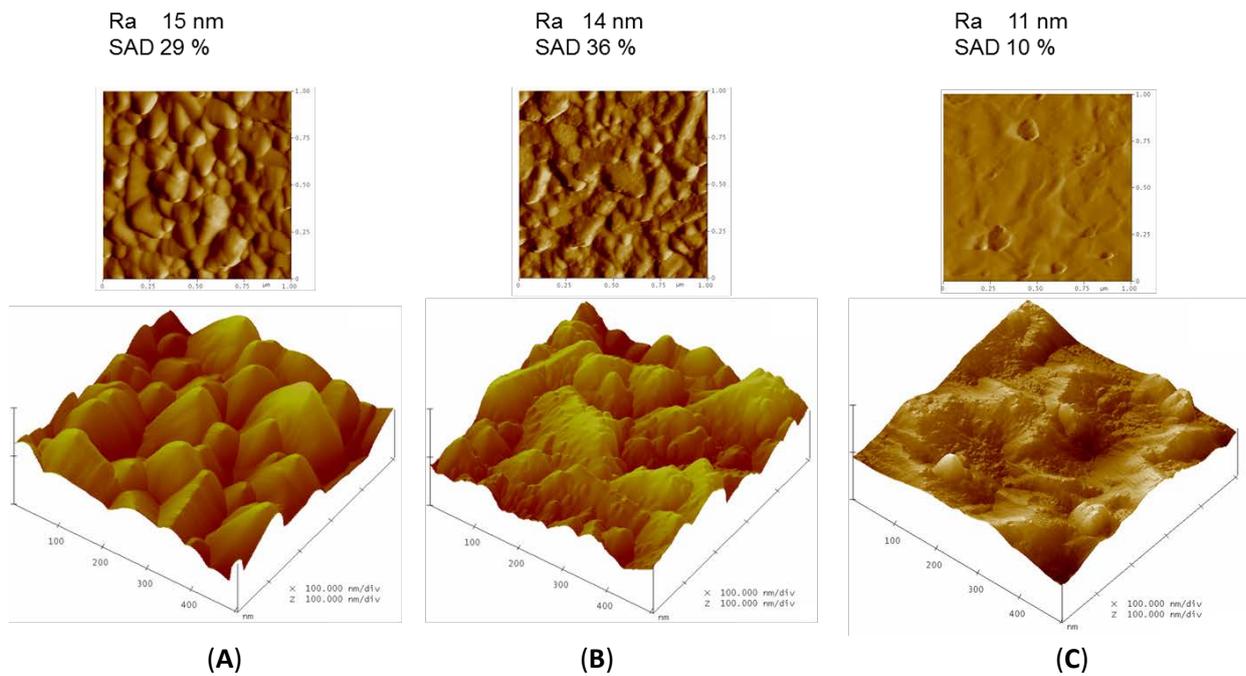

**Figure 1:** AFM images - tapping amplitude (top) and 3D topography/height (bottom) of the starting substrates, i.e. FTO as received (A) and that coated by Pt (B) and by PtGONF (C). While images of the as-received surface show sub-micro grains with smooth surface (A), platinum coating manifests itself as nanograins well resolved on the surface of larger sub-micrograins (B), PtGONF coating causes surface smoothening with still visible Pt nanoparticles (C)

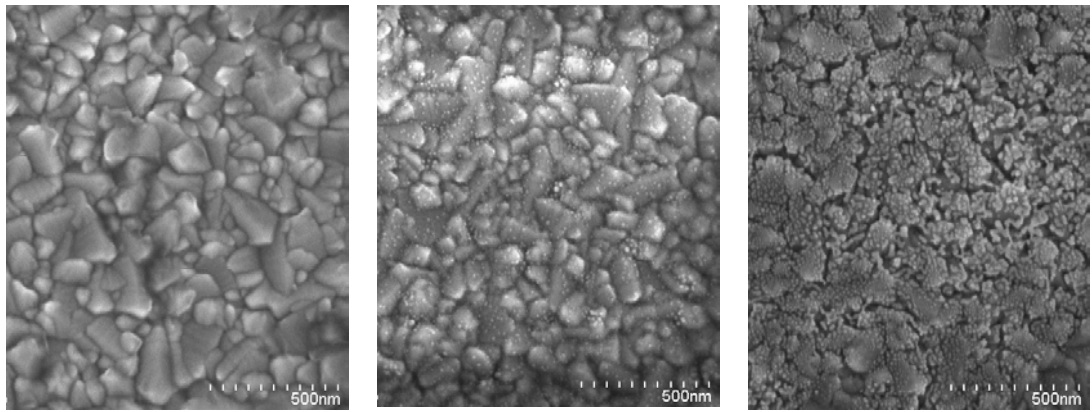

| A | B | C |

**Figure 2**: SEM image of FTO surface as received (A), coated by Pt (B) and by PtGONF (C).

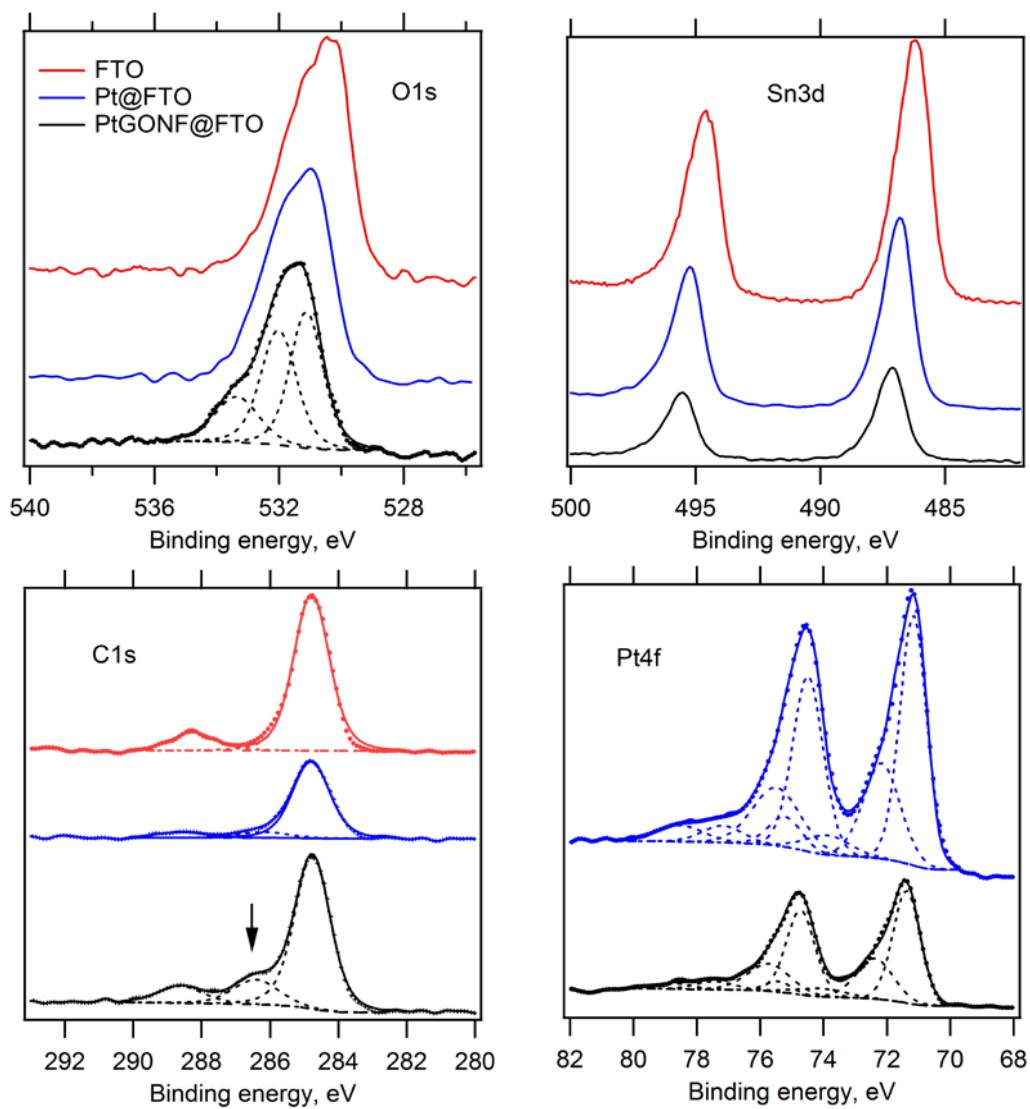

**Figure 3**: XPS spectra of FTO surface as received (red curves), coated by Pt (blue curves) and by PtGONF (black curves). Spectra are offset for clarity but the intensity scale is identical in each chart. Arrow points at the C1s line which is specific for graphene-oxide.

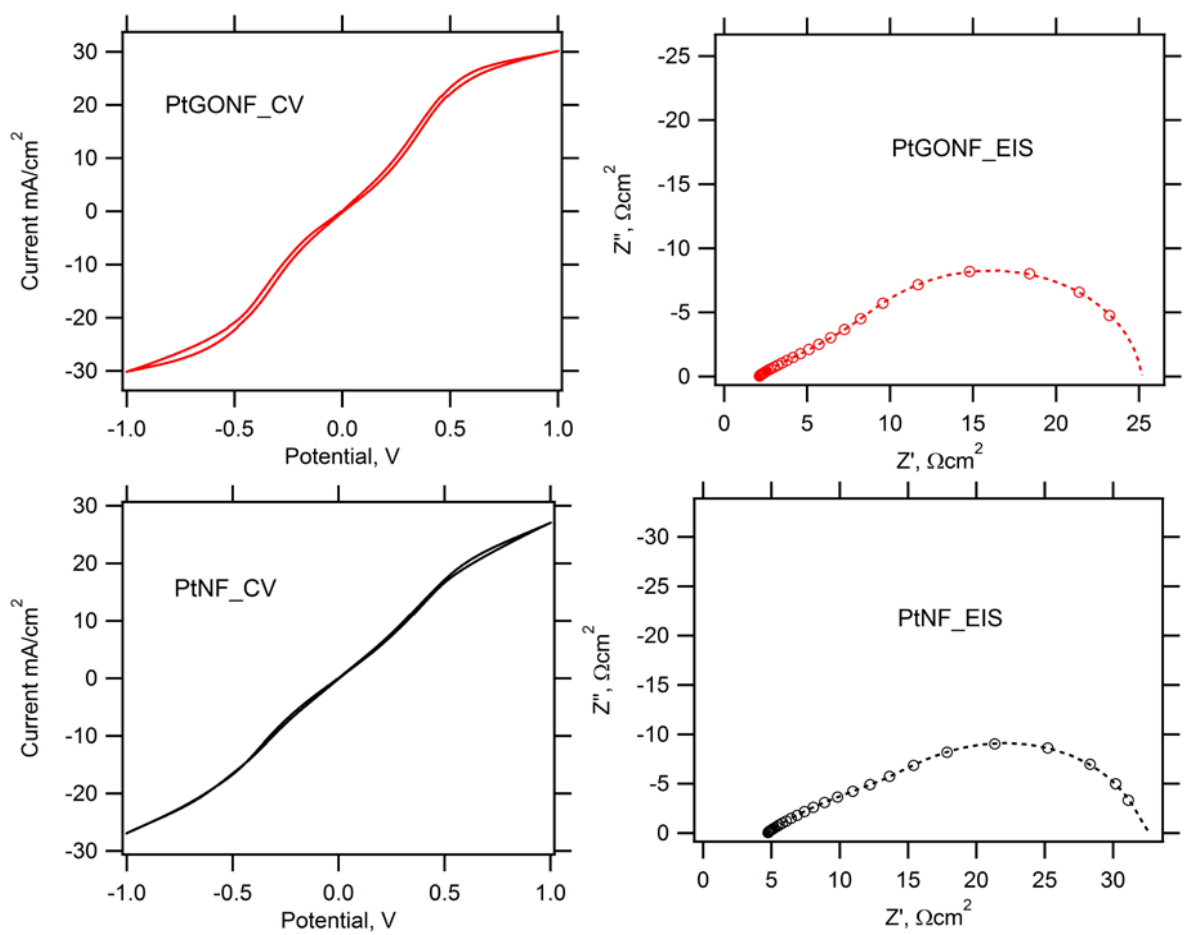

**Figure 4**: Electrochemical activity of the optimized FTO-supported catalytic electrodes tested on symmetrical dummy cells. Left charts: Cyclic voltammograms, scan rate 10 mV/s. Right charts: Nyquist plots of electrochemical impedance spectra measured at 0 V from 100 kHz to 0.1 Hz. (Markers are experimental points, dashed lines are simulated fits to the equivalent circuit). The catalysts are labeled PtGONF and PtNF, respectively; for details see text.

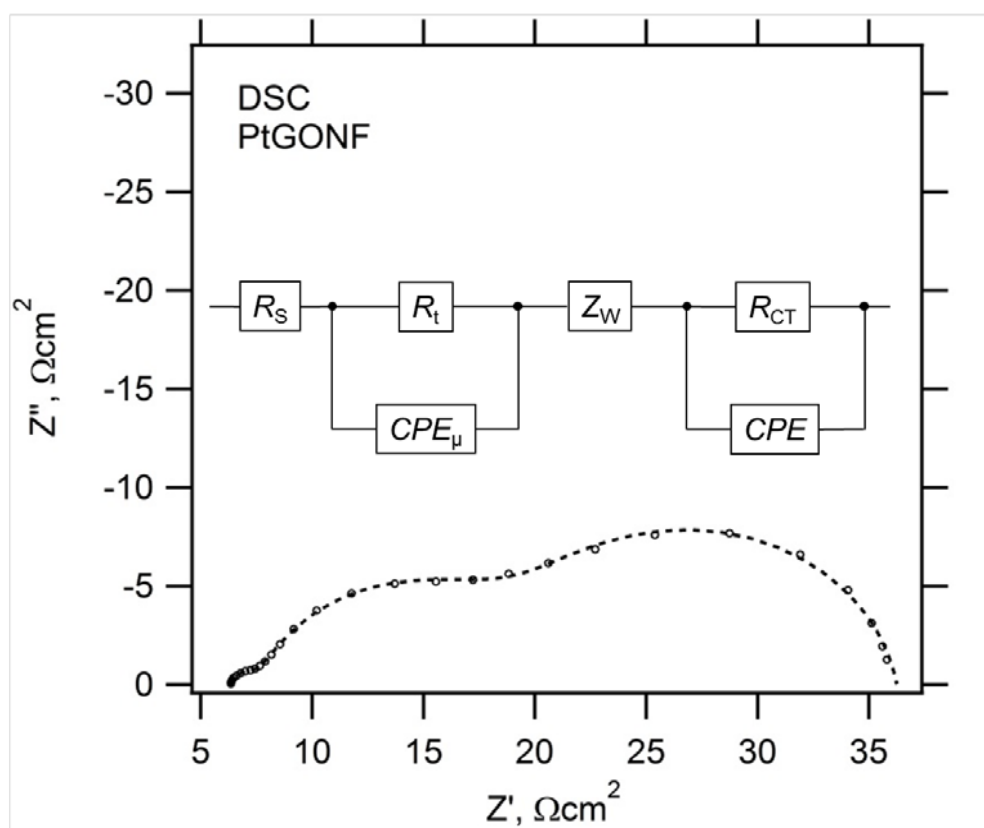

**Figure 5.** Electrochemical imedance spectrum of a solar cell with Y123-sensitized $TiO_2$ photoanode and PtGONF counterelectrode measured in dark at the applied forward bias of -1 V. Open circles are experimental points, dashed curve is a fitted spectrum. Inset shows the simplified equivalent circuit which was used for spectra fitting: $R_S$ = ohmic serial resistance, $Z_W$ = Warburg impedance in the electrolyte solution, $R_t$ = recombination resistance at the $TiO_2$/electrolyte solution interface, $CPE_\mu$ = constant phase element modeling the chemical capacitance of $TiO_2$, $R_{CT}$ = charge-transfer resistance at counterelectrode, $CPE$ = constant phase element modelling the counter-electrode.

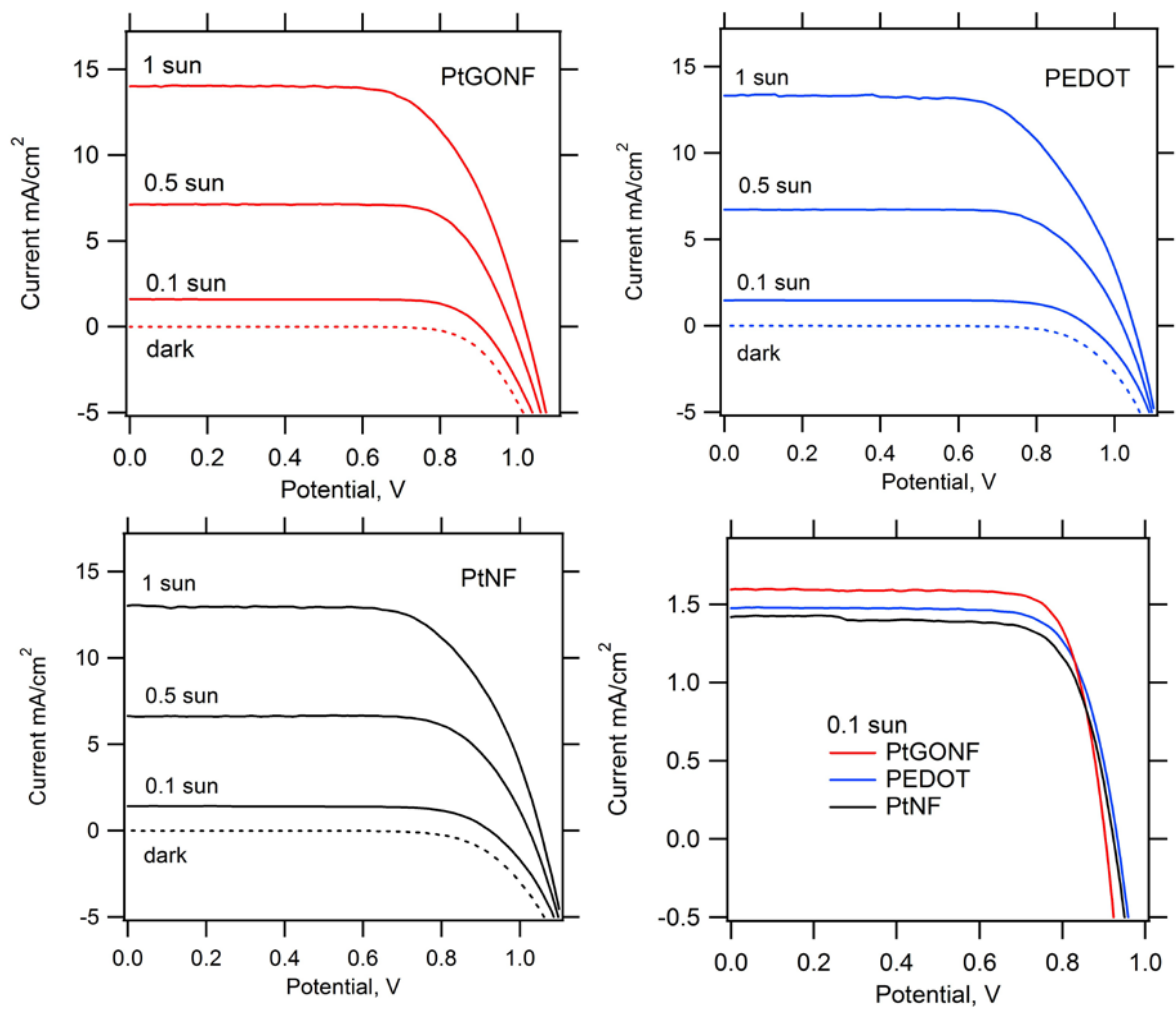

**Figure 6:** Current-voltage characteristics of dye-sensitized solar cells measured under standard AM1.5G illumination (1 sun) and at the proportionally attenuated light intensity (0.5 sun and 0.1 sun). The solar cells employed the Y123-senstizied $TiO_2$ photoanode and three different catalytic counterelectrodes, viz. PtGONF, PEDOT and PtNF. Also shown is the comparison of current/voltage plots for 0.1 sun illumination (right bottom chart).